\documentclass[10pt,pra,showpacs,superscriptaddress,twocolumn]{revtex4}
\usepackage{mathrsfs}
\usepackage{graphics,color,amsmath,amssymb,graphicx,amsfonts}
\usepackage{times}

\begin{document}

\title{Half-vortex sheets and domain-wall trains of rotating two-component
Bose-Einstein condensates in spin-dependent optical lattices}
\author{Wei Han}
\affiliation{Institute of Theoretical Physics, Shanxi University, Taiyuan 030006, China}
\affiliation{Beijing National Laboratory for Condensed Matter Physics, Institute of
Physics, Chinese Academy of Sciences, Beijing 100190, China}
\author{Suying Zhang}
\affiliation{Institute of Theoretical Physics, Shanxi University, Taiyuan 030006, China}
\author{Jingjing Jin}
\affiliation{Institute of Theoretical Physics, Shanxi University, Taiyuan 030006, China}
\author{W. M. Liu}
\affiliation{Beijing National Laboratory for Condensed Matter Physics, Institute of
Physics, Chinese Academy of Sciences, Beijing 100190, China}
\date{\today}

\begin{abstract}
We investigate half-vortex sheets and domain-wall trains of rotating
two-component Bose-Einstein condensates in spin-dependent optical lattices.
The two-component condensates undergo phase separation in the form of
stripes arranged alternatively. The vortices of one component are aligned in
lines in the low-density regions and filled with the other component, which
results in a stable vortex configuration, \textit{straight half-vortex sheets%
}. A train of novel domain walls, with spatially periodic
\textquotedblleft eyebrow-like" spin textures embedded on them, are
formed at the interfaces of the two components. We reveal that these
spatially periodic textures on the domain walls result from the
linear gradient of the relative phase, which is induced by the
alternating arrangement of the vortex sheets in the two components.
An accurate manipulation of the textures can be realized by
adjusting the intercomponent interaction strength, the rotating
angular frequency and the period of the optical lattices.
\end{abstract}

\maketitle

\section{Introduction}

In recent years, the study on topological defects has become a fascinating
topic in Bose-Einstein condensates (BECs). In a single-component BEC,
topological defects manifest themselves as integer vortices \cite%
{single-component,Freilich,Jezek,Wang}. Multi-component BECs, which are
described by a vector order parameter, allow the existence of more variety
of exotic topological defects, such as fractional vortices \cite%
{multi-component,Semenoff,Ji,Eto,Su}, domain walls \cite%
{multi-component,Sadler} and textures \cite%
{multi-component,Yi,Huhtamaki,Kawakami,Su2}. As the simplest example of the
multi-component condensates, two-component BECs have also attracted much
interest to study various topological defects.

There are various vortex configurations in rotating two-component BECs, such
as triangular vortex lattices, rectangular vortex lattices, vortex sheets,
rotating droplets and giant vortices \cite%
{Mueller,Kasamatsu,Schweikhard,Woo,Yang,Kasamatsu2,Mason}. Different
configurations depend on the intracomponent interaction, the intercomponent
interaction and the particle numbers. Especially, when the intercomponent
interaction is strong enough and the imbalance of the intracomponent
parameter is small, \textquotedblleft serpentine" vortex sheets can be
formed \cite{Kasamatsu,Kasamatsu2}. However, this configuration is
disordered and not well controlled. Although straight vortex sheets were
also discussed in the previous work \cite{Woo,Kasamatsu2}, stable ones have
never been obtained. This is because there exist different metastable vortex
sheet configurations with almost the same energy as the straight one.

Due to the spin degrees of freedom, the two-component BECs can be considered
as a novel magnetic material. When the condensates undergo phase separation,
domain walls are formed naturally at the interfaces of the two components.
There have been several studies of domain walls in two-component BECs \cite%
{Coen,Son,Garcia,Kevrekidis,Malomed,Takeuch,Jin}. Most of them concentrate
on the formation and dynamics of domain walls, while the internal structure
of domain walls has not been explored so much. In contrast, in common
magnetic materials, various internal structures of domain walls have been
extensively investigated both theoretically and experimentally (see, e.g.,
Ref. \cite{Lee} and references therein). In order to study domain walls in
two-component BECs, phase separation is required. Usually, phase separation
is realized by the strong intercomponent repulsion. As the experimental
realization of the spin-dependent optical lattices \cite{Mandel,McKay}, the
two-component condensates can be separated in an arbitrary form, which
provides us a well-controlled platform to study domain walls.

A variety of spin textures are discovered in two-component BECs, such as
skyrmions \cite{Battye}, meron-pairs \cite{Kasamatsu3} and spin-2 textures
\cite{Ruben}. The motifs of different textures always correspond to
different vortex configurations. For example, an axisymmetric vortex and a
nonaxisymmetric one correspond to a skyrmion and a meron-pair, respectively
\cite{Kasamatsu4}, and a spin-2 texture corresponds to a pair of vortices
with opposite sign that reside in different components \cite{Ruben}. This
suggests that it is a feasible method to produce novel textures by
controlling the arrangement of the vortices.

In this paper, we investigate topological defects of rotating
two-component BECs in spin-dependent optical lattices. We find that
this system supports a new stable vortex configuration, straight
half-vortex sheets. A train of novel domain walls are formed at the
interfaces of the two components. We concentrate on their unique
internal structures, and find that spatially periodic
\textquotedblleft eyebrow-like" spin textures are embedded on the
domain walls. We reveal that these spatially periodic textures are
directly determined by the arrangement of the straight half-vortex
sheets. The influences of the system parameters on the textures are
also investigated both analytically and numerically. Our results
show that the number of the textures on a wall is proportional to
the rotating angular frequency and the period of the optical
lattices, and with the increase of the intercomponent interaction
strength, the textures become thinner and thinner. This allows us to
make an accurate manipulation of the textures.

This paper is organized as follows. In Sec. \uppercase\expandafter{%
\romannumeral2}, we give the model of the rotating two-component BECs in
spin-dependent optical lattices. In Sec. \uppercase\expandafter{%
\romannumeral3}, we study the stable straight half-vortex sheets, and
discuss the discontinuity of the tangential component of the superfluid
velocity across the sheet. In Sec. \uppercase\expandafter{\romannumeral4},
we investigate the internal structures of the domain walls and reveal the
formation mechanism of the spatially periodic \textquotedblleft
eyebrow-like" spin textures. In Sec. \uppercase\expandafter{\romannumeral5},
we focus on the texture control. We conclude this paper in Sec. \uppercase%
\expandafter{\romannumeral6}.

\section{The Model}

We consider a two-level $^{\text{87}}$Rb BEC system with $\left\vert
F=1,m_{f}=-1\right\rangle \equiv \left\vert 1\right\rangle $ and $\left\vert
F=2,m_{f}=1\right\rangle \equiv \left\vert 2\right\rangle $ \cite{Mertes}.
In the weak interaction limit, the two-component condensates in a frame
rotating at an angular frequency $\Omega $ around the $z$ axis can be
described by the coupled Gross-Pitaevskii (GP) equations%
\begin{eqnarray}
i\hbar \frac{\partial \Psi _{i}\left( \mathbf{r},t\right) }{\partial t} &=&%
\Bigl[-\frac{\hbar ^{2}}{2m}\mathbf{\nabla }^{2}+V_{\text{H}%
}+\sum_{j=1,2}g_{ij}\left\vert \Psi _{j}\right\vert ^{2}  \notag \\
&&+V_{\text{OL}i}-\Omega \hat{L}_{z}\Bigr]\Psi _{i}\left( \mathbf{r}%
,t\right) ,  \label{3D GP equation}
\end{eqnarray}%
where $\Psi _{i}$ is the macroscopic wave function of the $i$th component ($%
i=1,2$). $g_{ij}=4\pi \hbar ^{2}a_{ij}/m$ represents the strength of
interatomic interactions characterized by the intra- and intercomponent $s$%
-wave scattering lengths $a_{ij}$ and the mass $m$ of an atom. $\hat{L}%
_{z}=-i\hbar \left( x\partial _{y}-y\partial _{x}\right) $ is the $z$%
-component of the angular momentum operator. The external potential consists
of two parts, the harmonic trapping potential $V_{\text{H}}=\frac{1}{2}m%
\left[ \omega _{\bot }^{2}\left( x^{2}+y^{2}\right) +\omega _{z}^{2}z^{2}%
\right] $ and the spin-dependent optical lattice potential $V_{\text{OL}i}$,
where $V_{\text{OL}1}=I_{0}\sin ^{2}(kx)$ and $V_{\text{OL}2}=I_{0}\cos
^{2}(kx)$. Here $k$ is the wave vector of the laser light used for the
optical lattice potentials and $I_{0}$ is the potential depth of the
lattices. The wave functions are normalized as $\sum_{i}\int \left\vert \Psi
_{i}\right\vert ^{2}d\mathbf{r}=N$, where $N$ is the total number of
condensate atoms.

For simplicity, we assume that the harmonic trapping frequencies satisfy $%
\omega _{z}\gg \omega _{\bot }$. Then, the condensates are pressed into a
\textquotedblleft pancake". This allows us to reduce Eq. (\ref{3D GP
equation}) to a two-dimensional form as \cite{Kasamatsu5}
\begin{eqnarray}
i\hbar \frac{\partial \psi _{i}\left( x,y,t\right) }{\partial t} &=&\Bigl[-%
\frac{\hbar ^{2}}{2m}\mathbf{\nabla }^{2}+\tilde{V}_{\text{H}%
}+\sum_{j=1,2}\eta g_{ij}\left\vert \psi _{j}\right\vert ^{2}  \notag \\
&&+V_{\text{OL}i}-\Omega \hat{L}_{z}\Bigr]\psi _{i}\left( x,y,t\right) ,
\label{2D GP equation}
\end{eqnarray}%
where $\eta =\left( h/m\omega _{z}\right) ^{-1/2}$ is a reductive parameter.
The two-dimensional wave functions are normalized as $\sum_{i}\int
\left\vert \psi _{i}\right\vert ^{2}dxdy=N$. The harmonic trapping potential
is reduced to its 2D form $\tilde{V}_{\text{H}}=\frac{1}{2}m\omega _{\bot
}^{2}\left( x^{2}+y^{2}\right) $, where the tilde will be omitted in the
following for simplicity.

In order to describe the system more clearly, the intuitive pictures of the
external potentials are presented in Figs. 1(a)--1(f). Experimentally, the
spin-dependent optical lattice potentials $V_{\text{OL}1}$ and $V_{\text{OL}%
2}$ can be realized by employing two counter-propagating blue-detuned laser
beams with the same frequency but perpendicular linear polarization vectors
\cite{Mandel}. A schematic of the spin-dependent optical lattices is
presented in Fig. 1(g).
\begin{figure}[tbp]
\centering \vspace{0cm} \hspace{0cm}\scalebox{0.56}{%
\includegraphics{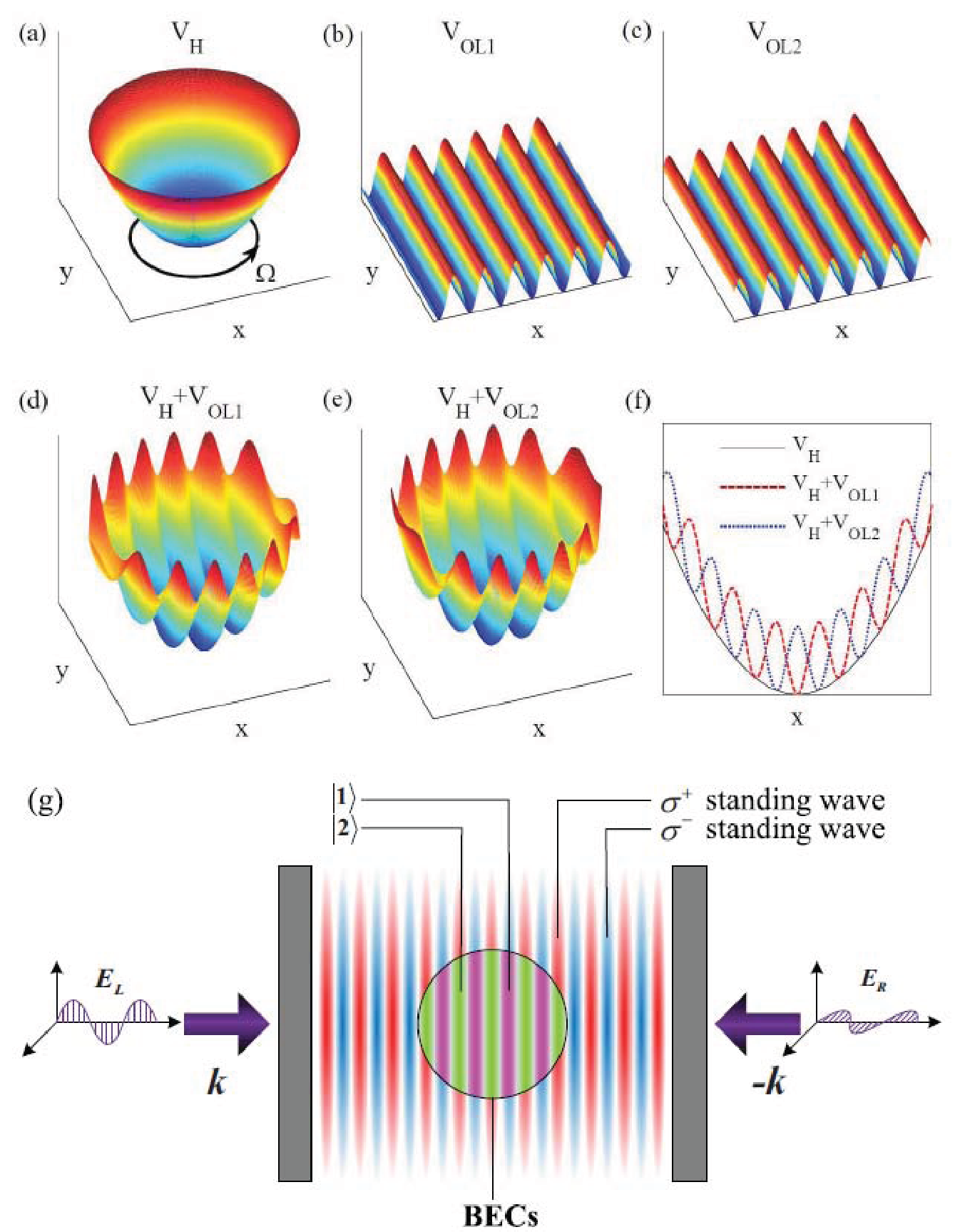}}
\caption{(Color online) (a) The harmonic trapping potential $V_{\text{H}}$
is rotated about the $z$ axis at a frequency $\Omega $. (b) The
spin-dependent optical lattice potential $V_{\text{OL}1}=I_{0}\sin
^{2}\left( kx\right) $, which is experienced by the $\left\vert
1\right\rangle $ state. (c) The spin-dependent optical lattice potential $V_{%
\text{OL}2}=I_{0}\cos ^{2}\left( kx\right) $, which is experienced by the $%
\left\vert 2\right\rangle $ state. (d) The composite potential $V_{\text{H}%
}+V_{\text{OL}1}$. (e) The composite potential $V_{\text{H}}+V_{\text{OL}2}$%
. (f) Cross sections of $V_{\text{H}}$ (solid line), $V_{\text{H}}+V_{\text{%
OL}1}$ (dashed line) and $V_{\text{H}}+V_{\text{OL}2}$ (dotted line) along
the $x$ axis. (g) Schematic of the spin-dependent optical lattices. Two
polarized standing wave laser fields $\protect\sigma ^{+}$ (red) and $%
\protect\sigma ^{-}$ (blue) are formed by two counter-propagating
blue-detuned laser beams with the same frequency but perpendicular linear
polarization vectors. This gives rise to the optical lattice potentials $V_{%
\text{OL}1}$ and $V_{\text{OL}2}$, which are experienced by the $\left\vert
1\right\rangle $ state (pink) and the $\left\vert 2\right\rangle $ state
(green), respectively. }
\label{fig1}
\end{figure}

In our simulations, the $^{\text{87}}$Rb atoms are assigned to the two
states equally and the total number of them is $N=10^{5}$. The radial and
axial trapping frequencies are $\omega _{\bot }=2\pi \times 15$ Hz and $%
\omega _{z}=2\pi \times 150$ Hz, respectively. We use the scattering lengths
\cite{Mertes}: $a_{11}=100.40a_{0},a_{22}=95.00a_{0}$ and $%
a_{12}=a_{21}=97.66a_{0}$ ($a_{0}$ is the Bohr radius), except when we
discuss the influence of the intercomponent interaction on the textures in
Sec. \uppercase\expandafter{\romannumeral5}. The intensity of the laser
light used for the optical lattice potentials is chosen as $%
I_{0}=k_{B}\times 27.5$ nK ($k_{B}$ is the Boltzmann's constant) \cite%
{Burger}, which is powerful enough such that the two states are phase
separated.

\section{Half-vortex sheets}

In this section, we present a stable vortex configuration, straight
half-vortex sheets. By using the imaginary-time propagation method \cite{Bao}%
, we solve Eq. (\ref{2D GP equation}) numerically and obtain the ground
state of the two-component condensates. The density profiles with the
rotating angular frequency $\Omega =0.6\omega _{\bot }$ and the period of
the optical lattice potential $T=\pi \xi $ are presented in Fig. 2, where $%
\xi =\left( \hbar /m\omega _{\bot }\right) ^{-1}$ is the\ spatial scale. All
the vortices are denoted by crosses ($\times $), whose positions are
determined by the singularities of the phase. From Fig. 2, we can see that
the two-component condensates undergo phase separation in the form of
stripes arranged alternatively. The vortices of one component are aligned in
lines in the low-density regions and filled with the other component. This
results in alternatively arranged straight vortex sheets in the two
components. Obviously, all the positions of the vortices in one component
are vortex-free regions in the other component, so all the vortices are half
quantized \cite{Ji,Eto,Su}. We refer to this vortex configuration as
straight half-vortex sheets. Even though the half-vortex sheets do not
influence the total density distribution of the condensates, they are
crucial for the formation of the spatially periodic spin textures on the
domain walls.

In the absence of the spin-dependent optical lattices, the straight vortex
sheets configuration has also been discussed in the phase-separated region
\cite{Woo,Kasamatsu2}. However, this configuration is ustable in that case,
and there exist many different shapes of metastable vortex sheet
configurations with almost the same energy as the straight one. This is
because the energy of the vortex sheets is mainly determined by the
intervortex spacing within a vortex sheet and the intersheet spacing rather
than the shape of the vortex sheets. In contrast, when the spin-dependent
optical lattices are present, the shape of the vortex sheets becomes an
important factor in determining the energy. Any bending in the straight
vortex sheets will cost much energy, so the straight half-vortex sheets
configuration obtained in our system is stable. In our imaginary-time
propagations, for any trial initial configurations, the straight vortex
sheets configuration is always uniquely obtained after sufficient
convergence of the energy.
\begin{figure}[tbp]
\centering \vspace{0cm} \hspace{0cm}\scalebox{0.34}{%
\includegraphics{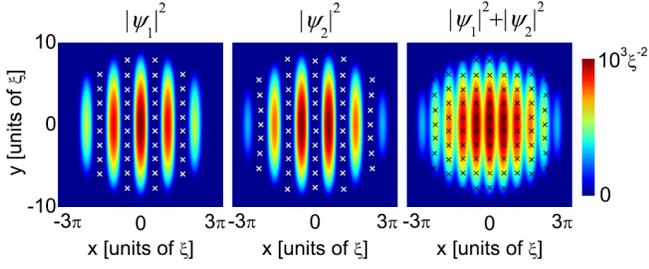}}
\caption{(Color online) The ground state density profiles of $\left\vert
\protect\psi _{1}\right\vert ^{2}$, $\left\vert \protect\psi _{2}\right\vert
^{2}$ and $\left\vert \protect\psi _{1}\right\vert ^{2}+\left\vert \protect%
\psi _{2}\right\vert ^{2}$ for the rotating angular frequency $\Omega =0.6%
\protect\omega _{\bot }$ and the period of the optical lattice potential $T=%
\protect\pi \protect\xi $ with $\protect\xi =\left( \hbar /m\protect\omega %
_{\bot }\right) ^{1/2}$. The locations of the vortices are marked by crosses
($\times $).}
\label{fig2}
\end{figure}

The most well-known character of a vortex sheet is the discontinuity of the
tangential component of the velocity across the sheet. The regular
arrangement of the straight vortex sheets allows us to observe this
phenomenon clearly. In Fig. 3(a), the tangential components (the components
along the $y$ axis) $v_{1y}$ and $v_{2y}$ of the superfluid velocities $%
\mathbf{v}_{1}$ and $\mathbf{v}_{2}$ \ of the two states are presented. We
can see that both $v_{1y}$ and $v_{2y}$ discontinuously jump across every
sheet. In order to describe the tangential velocities in detail, the section
views of $v_{1y}$ and $v_{2y}$ along the $x$ axis are shown in Fig. 3(b).
The $y$ component $v_{y}^{\text{rb}}=\Omega x$ of the rigid body rotation
velocity $\mathbf{v}^{\text{rb}}=\mathbf{\Omega }\times \mathbf{r}$ is
presented for comparison. We find that both the tangential velocities $v_{1y}
$ and $v_{2y}$ have a sawtoothlike change following the rigid-body value $%
v_{y}^{\text{rb}}$. The value of $v_{iy}$ jumps $2\Omega b$ across the sheet
in each component and then decreases $\Omega b$ linearly in a intersheet
spacing $b$. Here, the intersheet spacing $b$ is defined as the length
between two neighboring sheets in the same component.
\begin{figure}[t]
\centering \vspace{0cm} \hspace{0cm}\scalebox{0.45}{%
\includegraphics{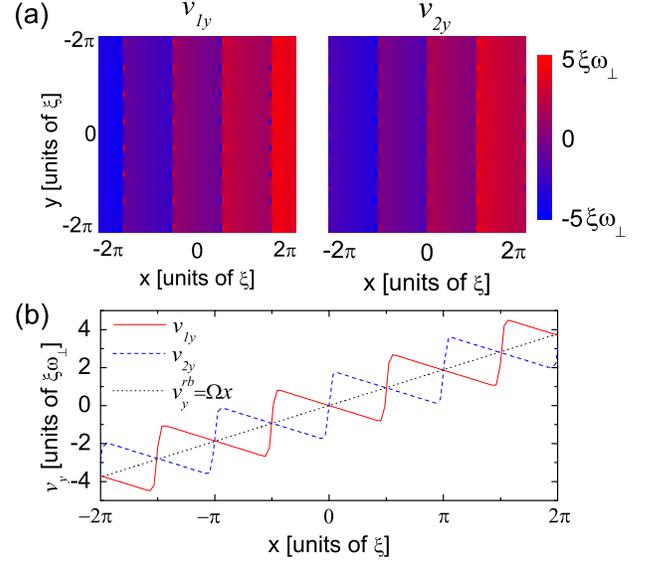}}
\caption{(Color online) (a) The tangential components $v_{1y}$ (left) and $%
v_{2y}$ (right) of the superfluid velocities $\mathbf{v}_{1}$ and $\mathbf{v}%
_{2}$ of the two states for $\Omega =0.6\protect\omega _{\bot }$ and $T=%
\protect\pi \protect\xi $. (b) Section views of $v_{1y}$ (solid line) and $%
v_{2y}$ (dashed line) along the $x$ axis. The $y$ component $v_{y}^{\text{rb}%
}$=$\Omega x$ of the rigid body rotation velocity $\mathbf{v}^{\text{rb}}$
is shown by the dotted line for comparison.}
\label{fig3}
\end{figure}
\begin{figure*}[tbp]
\centering \vspace{0cm} \hspace{0cm}\scalebox{0.95}{%
\includegraphics{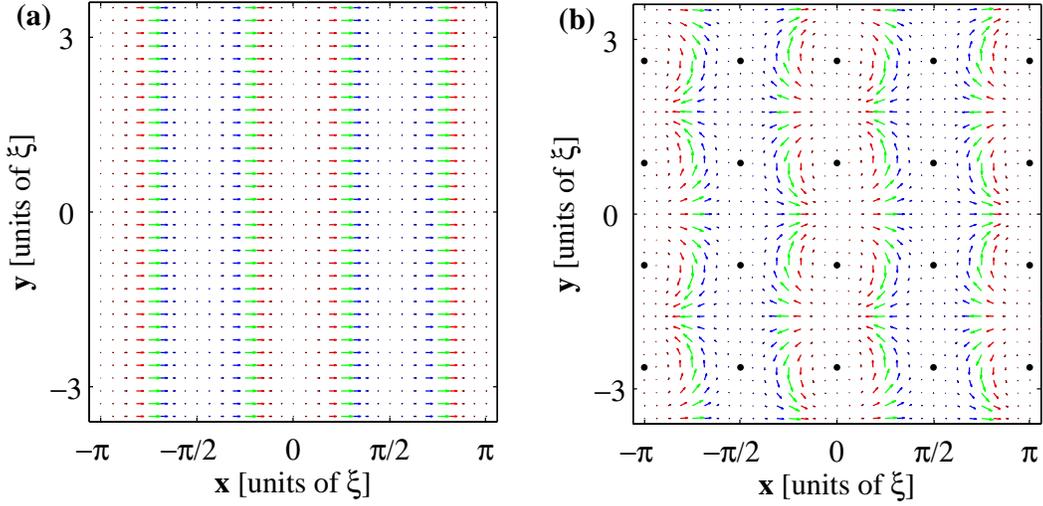}}
\caption{(Color online) The vectorial representations of the pseudospin $%
\mathbf{S}$ projected onto the $x$-$y$ plane for (a) $\Omega =0$, $T=\protect%
\pi \protect\xi $ and (b) $\Omega =0.6\protect\omega _{\bot }$, $T=\protect%
\pi \protect\xi $. The colors ranging from blue to red describe the values
of $S_{z}$ from $-1$ to $1$. The locations of the vortices are marked by
black dots ($\bullet $).}
\label{fig4}
\end{figure*}

The numerical results obtained above can be understood analytically.\emph{\ }%
Considering that the $y$ component of the rigid body rotation velocity is $%
v_{y}^{\text{rb}}=\Omega x$, we suppose that the tangential components of
the superfluid velocity on both sides of the sheet $v_{iy}^{-}$ and $%
v_{iy}^{+}$ are independent of $y$. According to Onsager-Feynman
quantization condition \cite{single-component}%
\begin{equation}
\oint\nolimits_{\mathcal{C}}\mathbf{v}_{\text{s}}\cdot d\mathbf{l}=\frac{%
2\pi \hbar }{m}N_{v},  \label{Feynman condition}
\end{equation}%
if we choose the two sides of the sheet as the integration path, we can
obtain that the tangential velocity jump across a sheet is
\begin{equation}
\Delta v_{iy}=v_{iy}^{+}-v_{iy}^{-}=\frac{2\pi \hbar }{m}\frac{1}{d_{v}},
\label{velocity jump 1}
\end{equation}%
where $d_{v}$ is the intervortex spacing within a vortex sheet. This implies
that the tangential velocity jump across the sheet is only determined by the
intervortex spacing within the sheet. As the mean vortex density of each
component can be estimated as
\begin{equation}
n_{1}=n_{2}=\frac{m\Omega }{\hbar \pi }\text{,}  \label{mean vortex density}
\end{equation}%
we can obtain that the intervortex spacing within a vortex sheet is%
\begin{equation}
d_{v}=\frac{\pi \hbar }{m\Omega T}\text{.}  \label{intervortex spacing}
\end{equation}%
Form Eq. (\ref{velocity jump 1}) and Eq. (\ref{intervortex spacing}), we have%
\begin{equation}
\Delta v_{iy}=2\Omega T.  \label{velocity jump 2}
\end{equation}%
As the intersheet spacing $b$ is just equal to the period of the optical
lattice potential $T$, the tangential velocity jump across the sheet can
also be expressed as
\begin{equation}
\Delta v_{iy}=2\Omega b\text{.}  \label{velocity jump 3}
\end{equation}%
Meanwhile, in order to follow the rigid-body value $v_{y}^{\text{rb}}$, $%
v_{iy}$ must decrease $\Omega b$ in a intersheet spacing. These analytical
results agree well with the numerical simulations above.

\section{Domain-wall trains}

The spinor order parameter of the two-component BECs allows us to analyze
this system as a pseudospin-1/2 BEC and take it as a magnetic system \cite%
{Kasamatsu4}. Introducing a normalized complex-valued spinor $\boldsymbol{%
\chi }$, we represent the two-component wave functions as $\psi _{i}=\sqrt{%
\rho _{T}\left( \mathbf{r}\right) }\chi _{i}\left( \mathbf{r}\right) $,
where $\rho _{T}\left( \mathbf{r}\right) $ is the total density and the
spinor satisfies $\left\vert \chi _{1}\right\vert ^{2}+\left\vert \chi
_{2}\right\vert ^{2}=1$. In pseudospin representation, the pseudospin
density is defined as $\mathbf{S}=\boldsymbol{\chi }^{\mathbf{T}}\boldsymbol{%
\sigma \chi }$, where $\boldsymbol{\sigma }$ is the Pauli matrix. Then we
have%
\begin{align}
S_{x}& =2\left\vert \chi _{1}\right\vert \left\vert \chi _{2}\right\vert
\cos \left( \theta _{1}-\theta _{2}\right) ,  \label{S_x} \\
S_{y}& =-2\left\vert \chi _{1}\right\vert \left\vert \chi _{2}\right\vert
\sin \left( \theta _{1}-\theta _{2}\right) ,  \label{S_y} \\
S_{z}& =\left\vert \chi _{1}\right\vert ^{2}-\left\vert \chi _{2}\right\vert
^{2},  \label{S_z}
\end{align}%
where $\theta _{i}$ is the phase of the wave function $\psi _{i}$.
\begin{figure}[tbp]
\centering \vspace{0cm} \hspace{0cm}\scalebox{0.32}{%
\includegraphics{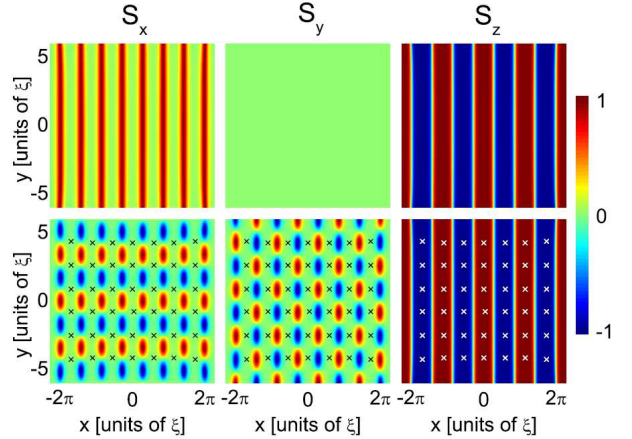}}
\caption{(Color online) The pseudospin densities $S_{x}$ (left), $S_{y}$
(middle), and $S_{z}$ (right). The upper panels show the case of $\Omega =0$
and $T=\protect\pi \protect\xi $, and the lower panels show the case of $%
\Omega =0.6\protect\omega _{\bot }$ and $T=\protect\pi \protect\xi $. The
locations of the vortices are marked by crosses ($\times $). }
\label{fig5}
\end{figure}

As the presence of the spin-dependent optical lattices, a train of domain
walls are formed naturally at the interfaces of the two components. By the
pseudospin representation, we investigate the response of the domain walls
to rotation. In order to reveal the essential influence of the rotation on
the structure of the domain walls, the non-rotating and rotating ground
states of the two-component condensates are calculated under the same
parameters. As the rotation can create an effective harmonic centrifugal
potential with frequency $\Omega $ \cite{Raman}, we change the radial
harmonic trapping frequency to $\sqrt{\omega _{\bot }^{2}-\Omega ^{2}}$ in
the absence of rotation. Thus, the non-rotating ground state density
distribution is nearly the same as the rotating ground state density
distribution, except that no vortex is created in the low-density regions of
each component in the non-rotating ground state.

The vectorial representation of the pseudospin $\mathbf{S}$ for the
non-rotating and rotating ground states are presented in Figs. 4(a) and
4(b), respectively. Correspondingly, the pseudospin densities $S_{x}$, $S_{y}
$ and $S_{z}$ are presented in Fig. 5. From Figs. 4 and 5, we find that a
train of domain walls are formed at the interfaces of the spin up ($S_{z}=1$%
) and spin down ($S_{z}=-1$) regions. In the absence of rotation, the
magnetic moments on the domain walls reverse only along the $x$ axis [see
Fig. 4(a)], and the pseudospin density $S_{y}=0$ [see the upper panels of
Fig. 5]. Therefore, these domain walls are classical N\'{e}el walls. In
contrast, in the presence of rotation, the magnetic moments on the domain
walls twist and form spatially periodic \textquotedblleft eyebrow-like" spin
textures [see Fig. 4(b) and the lower panels of Fig. 5]. This results in a
train of novel domain walls with spatially periodic textures embedded on
them.
\begin{figure}[tbp]
\centering \vspace{0cm} \hspace{0cm}\scalebox{0.87}{%
\includegraphics{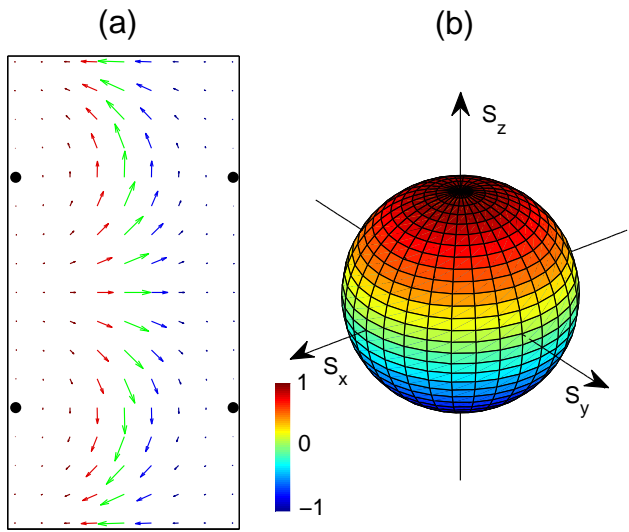}}
\caption{(Color online) (a) An amplification of Fig. 4(b) in one period. (b)
Bloch sphere of the pseudospin density vector $\mathbf{S}$. Values of $S_{z}$
are represented by linear levels from blue to red ($-1$ to $1$). }
\label{fig6}
\end{figure}

It should be indicated that the classical N\'{e}el wall does not carry
topological charges. However, the twist of the magnetic moments makes the
novel domain wall carry topological charges. According to the topological
charge density
\begin{equation}
q\left( \mathbf{r}\right) =\frac{1}{8\pi }\epsilon ^{ij}\mathbf{S}\cdot
\partial _{i}\mathbf{S}\times \partial _{j}\mathbf{S},
\label{topological charge density}
\end{equation}%
we can calculate that each period of the textures on the domain wall just
carries one unit topological charge. As shown in Figs. 4 and 5, each pair of
adjacent vortices from the two components just induce half texture. Since a
vortex in one component is adjacent to two vortices in the other component,
on average, one vortex indeed induces one half unit of topological charge.
Therefore, the number of the vortices is just half the number of the
topological charges.

Next, we fasten our attention on the structure of the \textquotedblleft
eyebrow-like" spin textures. An amplification of Fig. 4(b) for one period is
presented in Fig. 6(a). From Eq. (\ref{S_x}) and Eq. (\ref{S_y}), the
direction that the magnetic moments reverses along just depends on the
relative phase, and can be represented by an azimuthal angle%
\begin{equation}
\alpha =\theta _{2}-\theta _{1}\text{.}  \label{azimuthal angle 1}
\end{equation}%
From Fig. 6(a), we can see that the azimuthal angle of the magnetic moments
changes from $-\pi $ to $\pi $ along the domain wall in a period, but it is
constant along the normal direction of the domain wall.

It is instructive to project the pseudospin density vector $\mathbf{S}$ onto
the surface of a unit Bloch sphere [see Fig. 6(b)]. Topologically, the
topological charge counts the times that the Bloch sphere are covered. So
one texture just covers the Bloch sphere once. From Fig. 6, we find that
walking through the domain wall from one side to the other corresponds to
strolling along a longitude line of the Bloch sphere from one pole to the
other, and walking along the domain wall corresponds to strolling along a
latitude line of the Bloch sphere. This is different from the case of a
skymion, for which walking along the radial direction of the skymion
corresponds to strolling along the longitude line of the Bloch sphere, and
walking along the azimuthal direction of the skymion corresponds to
strolling along the latitude line of the Bloch sphere. Therefore, this new
texture essentially corresponds to a skyrmion in the polar coordinates
instead of the Cartesian ones.
\begin{figure}[tbp]
\centering \vspace{0cm} \hspace{0cm}\scalebox{0.37}{%
\includegraphics{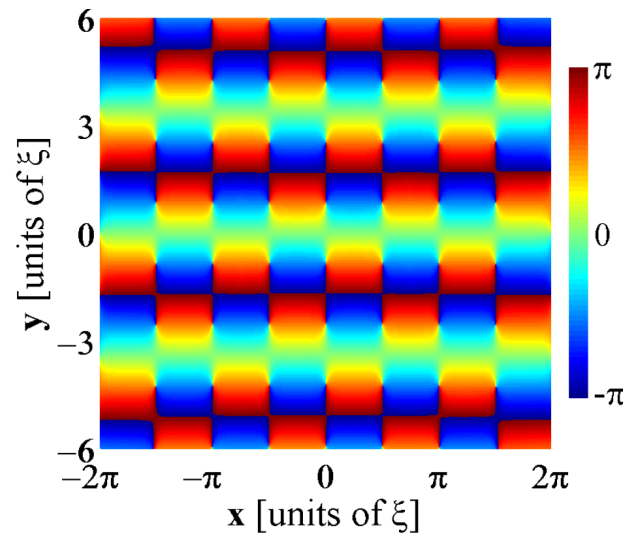}}
\caption{(Color online) The relative phase $\protect\theta _{1}-\protect%
\theta _{2}$ for $\Omega =0.6\protect\omega _{\bot }$ and $T=\protect\pi
\protect\xi $, where $\protect\theta _{1}$ and $\protect\theta _{2}$ are the
phases of the wave functions $\protect\psi _{1}$ and $\protect\psi _{2}$,
respectively.}
\label{fig7}
\end{figure}

We reveal the formation mechanism of the spatially periodic
\textquotedblleft eyebrow-like" spin textures on the domain walls. In the
absence of rotation, there is no relative phase between the two components
and the azimuthal angle $\alpha =0$. The magnetic moments on the domain
walls only reverse along the $x$ axis. Therefore, the domain walls are
classical N\'{e}el walls. In the presence of rotation, straight vortex
sheets are created in the two components and arranged alternatively on two
sides of the domain walls. These alternatively arranged vortex sheets induce
a linear gradient of the relative phase along the domain walls [see Fig. 7].
So the azimuthal angle $\alpha $ can be expressed as%
\begin{equation}
\alpha =\mathcal{P}\left( \kappa y\right) ,  \label{azimuthal angle 2}
\end{equation}%
where $\mathcal{P}$ projects the angle $\kappa y$ onto $\left( -\pi ,\pi %
\right] $ and $\kappa $ is a constant coefficient, which describes the
spatial change frequency of the azimuthal angle. The value of $\kappa $ will
be given in the next section. From Eq. (\ref{azimuthal angle 2}), the
azimuthal angle of the magnetic moments on the domain walls changes
periodically along the $y$ direction, and spatially periodic
\textquotedblleft eyebrow-like" spin textures are formed. This suggests that
the spatially periodic \textquotedblleft eyebrow-like" spin textures on the
domain walls result from the linear gradient of the relative phase, which is
induced by the alternating arrangement of the straight vortex sheets in the
two components.

\section{Texture control}

\begin{figure}[tbp]
\centering \vspace{0cm} \hspace{0cm}\scalebox{0.332}{%
\includegraphics{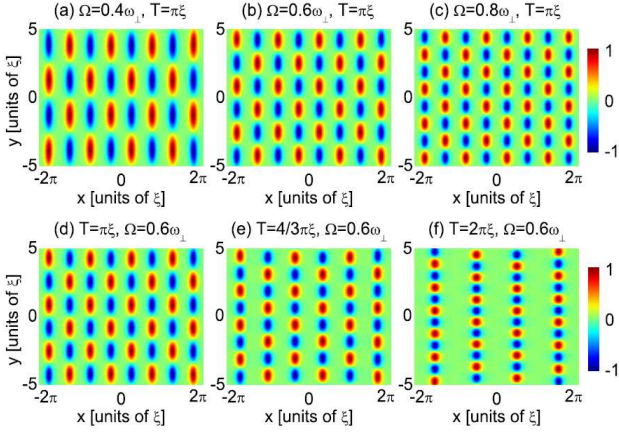}}
\caption{(Color online) The pseudospin density $S_{y}$ for different
rotating angular frequencies $\Omega $ and periods of the optical lattice
potential $T$. The upper panels show $S_{y}$ for $T=\protect\pi \protect\xi $
with (a) $\Omega =0.4\protect\omega _{\bot }$, (b) $\Omega =0.6\protect%
\omega _{\bot }$ and (c) $\Omega =0.8\protect\omega _{\bot }$. The lower
panels show $S_{y}$ for $\Omega =0.6\protect\omega _{\bot }$ with (d) $T=%
\protect\pi \protect\xi $, (e) $T=4/3\protect\pi \protect\xi $ and (f) $T=2%
\protect\pi \protect\xi $.}
\label{fig8}
\end{figure}
In this section, we discuss how to control the \textquotedblleft
eyebrow-like" textures on the domain walls. Firstly, we study the
distribution of the textures analytically. The topological charge density $q$
has another formulation derived from the effective velocity \cite{Mueller2}
\begin{equation}
q\left( \mathbf{r}\right) =\frac{m}{2\pi \hbar }\left( \nabla \times \mathbf{%
v}_{\text{eff}}\right) _{z},  \label{topological charge density 2}
\end{equation}%
where the effective velocity is defined as%
\begin{equation}
\mathbf{v}_{\text{eff}}=\frac{\left( \rho _{1}\mathbf{v}_{1}+\rho _{2}%
\mathbf{v}_{2}\right) }{\rho _{T}},  \label{effective velocity}
\end{equation}%
with $\rho _{i}$\ the density of each component. Approximately treating the
effective velocity $\mathbf{v}_{\text{eff}}$ as the classical rigid body
value $\mathbf{v}_{\text{rb}}=\mathbf{\Omega }\times \mathbf{r}$, we obtain
the mean topological charge density
\begin{equation}
\bar{q}=\frac{m\Omega }{\pi \hbar }.  \label{mean topological charge density}
\end{equation}%
From Eq. (\ref{mean topological charge density}), we can calculate that the
topological charge on a domain wall per unit length is
\begin{equation}
\eta _{q}=\frac{m\Omega T}{2\pi \hbar }.
\label{topological charge per unit length}
\end{equation}%
This implies that the number of the textures carried by a domain wall is
proportional to the rotating angular frequency $\Omega $ and the period of
the optical lattice potential $T$. From Eq. (\ref{topological charge per
unit length}), the spatial change frequency $\kappa $ of the azimuthal angle
$\alpha $ in Eq. (\ref{azimuthal angle 2}) can be calculated as%
\begin{equation}
\kappa =2\pi \eta _{q}=\frac{m\Omega T}{\hbar }.
\label{spatial change frequency}
\end{equation}%
Thus, we obtain the azimuthal angle%
\begin{equation}
\alpha =\mathcal{P}\left( \frac{m\Omega T}{\hbar }y\right) .
\label{azimuthal angle 3}
\end{equation}%
\begin{figure}[tbp]
\centering \vspace{0cm} \hspace{0cm}\scalebox{0.47}{%
\includegraphics{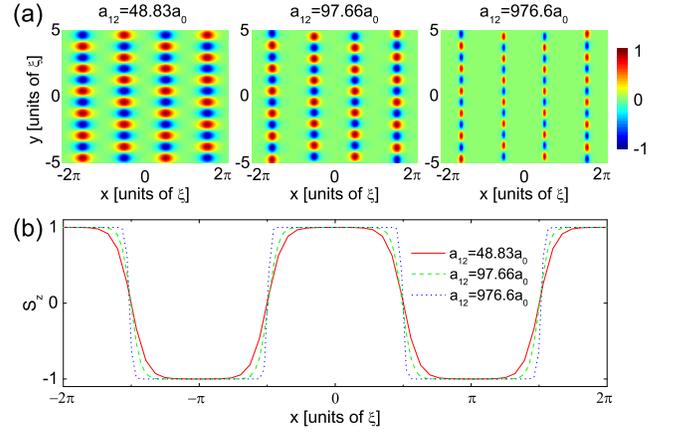}}
\caption{(Color online) (a) The pseudospin density $S_{y}$ for the
intercomponent scattering length $a_{12}=48.83a_{0}$ (left), $97.66a_{0}$
(middle), and $976.6a_{0}$ (right) with the intracomponent scattering
lengths $a_{11}=100.40a_{0}$ and $a_{22}=95.00a_{0}$, $\Omega =0.6\protect%
\omega _{\bot }$ and $T=2\protect\pi \protect\xi $. (b) Cross section views
of $S_{z}$ along the $x$ axis for $a_{12}=48.83a_{0}$ (solid line), $%
97.66a_{0}$ (dashed line), and $976.6a_{0}$ (dotted line).}
\label{fig9}
\end{figure}

In order to verify the above analytical discussion, we perform numerical
simulations. The pseudospin density $S_{y}$ for different rotating angular
frequencies $\Omega $ with constant period of the optical lattice potential $%
T$ is shown in the upper panels of Fig. 8, and $S_{y}$ for different $T$
with constant $\Omega $ is shown in the lower panels. Obviously, the number
of the topological charges carried by a domain wall increases in direct
proportion with the increase of $\Omega $ and $T$. For quantitative
comparison, we choose $\Omega =0.6\omega _{\bot }$ and $T=\pi \xi $ as an
example. From Eq. (\ref{topological charge per unit length}), we can
calculate that the number of the textures on a domain wall in the region of $%
y=[-5\xi ,5\xi ]$ is $3$. This agrees well with the result of the numerical
simulation in Fig. 8(b).

In our system, as the presence of the spin-dependent optical lattices, the
two components are always phase separated and not subject to the immiscible
condition, $g_{12}^{2}>g_{11}g_{22}$ \cite{Timmermans}. However, the
intercomponent interaction $g_{12}$ plays an important role on the domain
wall width. As the textures are always concentrated on the walls, we can
control the width of the textures by adjusting the intercomponent
interaction $g_{12}$. The pseudospin densities $S_{y}$ for different
intercomponent scattering length $a_{12}$ with constant intracomponent
scattering lengths $a_{11}$ and $a_{22}$ are shown in Fig. 9(a). The cross
section view of $S_{z}$ along the $x$ axis is shown in Fig. 9(b). From Fig.
9, we can see that with increasing the strength of intercomponent
interaction, the domain walls become narrower and narrower and the textures
on the walls become thinner and thinner.

\section{Conclusions}

We have investigated half-vortex sheets and domain-wall trains of rotating
two-component BECs in spin-dependent optical lattices. A stable vortex
configuration named straight half-vortex sheets is obtained. The
discontinuity of the tangential component of the superfluid velocity across
the sheet is discussed both numerically and analytically. We have also
investigated the response of the domain walls to rotation. In the absence of
rotation, the domain walls are classical N\'{e}el walls with the magnetic
moments only reversing perpendicular to the walls. In response to rotation,
the magnetic moments on the domain walls twist and form spatially periodic
\textquotedblleft eyebrow-like\textquotedblright\ spin textures. We have
revealed that these spatially periodic textures are directly determined by
the arrangement of the straight half-vortex sheets. The number of the
textures carried by the domain walls can be accurately controlled by
adjusting the rotating angular frequency and the period of the optical
lattice, and the width of the textures can be controlled by adjusting the
strength of intercomponent interaction. This allow us to make an accurate
manipulation of the \textquotedblleft eyebrow-like\textquotedblright\
textures.

With the development of the magnetization-sensitive phase-contrast imaging
technique \cite{Higbie}, both the longitudinal and transverse magnetization
of the domain and domain wall in BECs can be imaged non-destructively with
high spatial resolution \cite{Sadler,Vengalattore}. We expect that the novel
domain walls with spatially periodic \textquotedblleft
eyebrow-like\textquotedblright\ spin textures would be observed in the
future experiments.

\section*{ACKNOWLEDGMENTS}

We are grateful to D.-S. Wang and S.-W. Song for stimulating discussions and
valuable suggestions. This work was supported by NSFC under grants Nos.
10934010, 10972125, 60978019, the NKBRSFC under grants Nos. 2009CB930701,
2010CB922904, 2011CB921502, 2012CB821300, NSFC-RGC under grants Nos.
11061160490 and 1386-N-HKU748/10, NSFSP under grants Nos. 2010011001-2, and
SFRSP.

\end{document}